\documentclass{article}

\usepackage[main, final]{neurips_2026}

\usepackage[utf8]{inputenc}
\usepackage[T1]{fontenc}
\usepackage{hyperref}
\usepackage{url}
\usepackage{booktabs}
\usepackage{amsfonts}
\usepackage{amsmath,amssymb,amsthm,mathtools}
\usepackage{nicefrac}
\usepackage{microtype}
\usepackage{xcolor}
\usepackage{float}
\usepackage{enumitem}
\usepackage{tikz}
\usepackage{tcolorbox}
\usetikzlibrary{arrows.meta,calc,positioning}
\hypersetup{hidelinks,hypertexnames=false}
\usepackage{subcaption}
\usepackage{tabularx}

\usepackage{aliascnt}

\newtheorem{theorem}{Theorem}[section]

\newaliascnt{lemma}{theorem}
\newtheorem{lemma}[lemma]{Lemma}
\aliascntresetthe{lemma}

\newaliascnt{proposition}{theorem}
\newtheorem{proposition}[proposition]{Proposition}
\aliascntresetthe{proposition}

\newaliascnt{corollary}{theorem}
\newtheorem{corollary}[corollary]{Corollary}
\aliascntresetthe{corollary}

\newaliascnt{definition}{theorem}
\newtheorem{definition}[definition]{Definition}
\aliascntresetthe{definition}

\newaliascnt{assumption}{theorem}

\aliascntresetthe{assumption}

\newaliascnt{remark}{theorem}
\newtheorem{remark}[remark]{Remark}
\aliascntresetthe{remark}

\newaliascnt{example}{theorem}

\aliascntresetthe{example}

\usepackage[capitalize,noabbrev]{cleveref}

\crefname{corollary}{Corollary}{corollaries}
\Crefname{corollary}{Corollary}{Corollaries}

\usepackage{tikz}
\usepackage{pgfplots}
\pgfplotsset{compat=1.18}

\definecolor{myOrange}{RGB}{235, 105, 92}
\definecolor{myMaroon}{RGB}{128, 0, 0}
\definecolor{maroon}{rgb}{0.5,0.0,0.0}
\definecolor{heavymaroon}{rgb}{0.3,0.0,0.0}

\definecolor{mygreen}{RGB}{160, 200, 140}

\hypersetup{ 
colorlinks=true, 
linkcolor=myOrange, 
citecolor=myMaroon, filecolor=cyan, 
urlcolor=mygreen
}

\usepackage{lipsum}                     %
\usepackage{xargs}
\usepackage{xfrac}
\usepackage{amssymb}
\usepackage{amsmath}
\usepackage{bm} 
\usepackage{mathtools} 
\usepackage{amsthm}
\usepackage{mleftright}
\usepackage{refcount}

\def\[#1\]{\begin{align*}#1\end{align*}}

\NewDocumentCommand{\numberthis}{om}{%
  \IfNoValueTF{#1}{%
    \refstepcounter{equation}\tag{\theequation}%
  }{%
    \tag{#1}%
  }%
  \label{#2}%
}

\newcommand{\R}{\mathbb{R}}
\newcommand{\Z}{\mathbb{Z}}
\newcommand{\bbF}{\mathbb{F}}

\newcommand{\FONE}{\operatorname{FONE}}
\newcommand{\AFONE}{\operatorname{AFONE}}
\newcommand{\PStat}{\operatorname{PStat}}
\newcommand{\Proj}{\operatorname{Proj}}

\newcommand{\calC}{\mathcal{C}}

\newcommand{\calR}{\mathcal{R}}

\newcommand{\calU}{\mathcal{U}}

\newcommand{\eps}{\varepsilon}

\newtheorem*{repeatthminner}{\repeatthmname}
\newcommand{\repeatthmname}{}

\newenvironment{repeatthm}[1]{%
  \renewcommand{\repeatthmname}{Theorem~\ref{#1}}%
  \begin{repeatthminner}%
}{%
  \end{repeatthminner}%
}

\newtheorem*{repeatlemmainner}{\repeatlemmaname}
\newcommand{\repeatlemmaname}{}

\newenvironment{repeatlemma}[1]{%
  \renewcommand{\repeatlemmaname}{Lemma~\ref{#1}}%
  \begin{repeatlemmainner}%
}{%
  \end{repeatlemmainner}%
}

\newtheorem*{repeatpropinner}{\repeatpropname}
\newcommand{\repeatpropname}{}

\newenvironment{repeatprop}[1]{%
  \renewcommand{\repeatpropname}{Proposition~\ref{#1}}%
  \begin{repeatpropinner}%
}{%
  \end{repeatpropinner}%
}

\newtheorem*{repeatcorollaryinner}{\repeatcorollaryname}
\newcommand{\repeatcorollaryname}{}

\newenvironment{repeatcorollary}[1]{%
  \renewcommand{\repeatcorollaryname}{Corollary~\ref{#1}}%
  \begin{repeatcorollaryinner}%
}{%
  \end{repeatcorollaryinner}%
}

\title{Homological Barriers to Stable Learning of First-Order Nash Equilibria}

\newcommand{\NeurIPSAuthorEntry}[3]{%
  \begin{tabular}[c]{@{}c@{}}
    {\normalfont\bfseries #1}\\[0.12em]
    {\normalfont #2}\\[0.12em]
    {\normalfont\ttfamily #3}
  \end{tabular}%
}

\author{%
  \begin{tabular}{@{}c@{\hspace{1em}}c@{}}
    \NeurIPSAuthorEntry
      {Ashkan Soleymani}
      {Massachusetts Institute of Technology}
      {ashkanso@mit.edu}
    &
    \NeurIPSAuthorEntry
      {Gabriele Farina}
      {Massachusetts Institute of Technology}
      {gfarina@mit.edu}
    \\[2.5em]
    \NeurIPSAuthorEntry
      {Patrick Jaillet}
      {Massachusetts Institute of Technology}
      {jaillet@mit.edu}
    &
    \NeurIPSAuthorEntry
      {Georgios Piliouras}
      {Google Deepmind}
      {gpil@google.com}
  \end{tabular}%
}

\begin{document}

\maketitle

\begin{abstract}
We identify a homological barrier to stable learning of local Nash equilibria, also known as first-order Nash equilibria (FONE), in constrained non-concave differentiable games. We consider continuous deterministic learning dynamics that fix every FONE point and aim to attract all initial conditions to the full FONE set. We show that this convergence guarantee is impossible when the target set has nontrivial positive-degree cohomology, such as that arising from a loop, and the state space is compact and contractible. We exhibit this barrier in a simple non-concave quadratic zero-sum game on a box whose exact FONE set consists of a diagonal six-edge loop and an isolated point. Consequently, no such dynamics can make the full FONE set a stable global attractor. We further show that degree two is the minimal polynomial degree at which this barrier can arise and extend the construction to obtain nontrivial homology in arbitrarily high degrees. Finally, we establish that the barrier persists for sufficiently small approximate-FONE targets under small perturbations, yielding an open, positive-volume family of quadratic zero-sum games with the same homological barrier.
\end{abstract}

\section{Introduction}

First-order equilibrium conditions are the workhorse of learning in non-concave games. When global Nash equilibria are difficult to compute or may fail to exist, the standard fallback is to seek a first-order Nash equilibrium (FONE), where each player satisfies the first-order optimality conditions for feasible unilateral deviations. In differentiable games with constrained strategy sets, this means that no player can improve their payoff to first order along a feasible direction. Although this condition is weaker than requiring each player to be a global best response, it provides a natural solution concept for studying the behavior and convergence of learning dynamics. FONE and related first-order stationarity notions are widely studied as solution targets and algorithmic benchmarks in the min-max and non-concave game literature \citep{daskalakis2018limit,adolphs2019local,nouiehed2019solving,mazumdar2020gradient,daskalakis2021complexity,ostrovskii2021efficient,daskalakis2022non,rockafellar2024generalized,bernasconi2024role,bernasconi2026complexity}.

We identify a topological barrier to learning FONE. The barrier is not algorithm-specific or computational; it arises from the topology of the target set. If the FONE set has nontrivial first homology, as in the presence of a nontrivial loop, then no continuous learning rule can both fix every FONE point and make the full FONE set a stable global attractor on a compact contractible state space. In our construction, projection is the source of this phenomenon. On the boundary of a box, normal-cone conditions permit outward-pointing gradients and can create cycles of projected stationary points, even for affine fields. We realize such a cycle in an explicit quadratic zero-sum game on a box. The resulting FONE set consists of a diagonal boundary cycle and an isolated point; the cycle creates the topological barrier, while the isolated point contributes no first homology. Thus, even simple non-concave quadratic games can defeat the usual pointwise-stationary global-convergence template. Our findings are in line the ``game dynamics as the meaning of a game'' perspective~\citep{papadimitriou2019game}, showing that understanding the long-run behavior of learning dynamics requires looking beyond the existence of equilibrium solutions.

Our contributions can be summarized as follows.
\begin{itemize}[leftmargin=1em, itemsep=2pt, topsep=1pt, parsep=2pt, partopsep=2pt]
    \item We formulate a general homological barrier to stable learning of FONE. Specifically, a compact target with nontrivial positive-degree \v{C}ech cohomology cannot be fixed pointwise and uniformly attracting under continuous dynamics on a compact contractible state space. This barrier applies to autonomous maps, time-varying deterministic maps, continuous-time semiflows through their time maps, and finite-memory methods with compact state spaces.

    \item We construct a quadratic zero-sum game on $[0,1]^3\times[0,1]^3$ whose exact FONE set consists of a diagonal six-edge cycle and an isolated point. The cycle gives the FONE set nontrivial first cohomology, and this triggers the barrier.

    \item 
    We prove that affine zero-sum games on boxes have contractible FONE sets, making degree two the minimum polynomial degree at which the barrier can arise. We further construct quadratic games whose FONE sets have nontrivial homology in arbitrarily high degrees. Finally, we show that the barrier remains under sufficiently small approximate-FONE targets and small $C^1$ perturbations, and identify an open, positive-volume family of quadratic zero-sum games with this property.
\end{itemize}

\section{Related Work}

Classical work on attractor stability and shape theory relates the topology of attracting sets to that of the ambient state space \citep{la1976stability,hastings1978shape,garay1991strong,kapitanski2000shape,bhatia2002stability}. In particular, a compact set that is fixed pointwise and uniformly attracting under continuous dynamics on a compact contractible state space must have trivial positive-degree \v{C}ech cohomology. This restriction provides the topological foundation for our analysis of stable learning dynamics and FONE sets in constrained games.

The closest game-dynamics antecedent is the Conley-theoretic impossibility theorem of \citet{milionis2023impossibility}. Their barrier arises from a \emph{degenerate finite game} whose mixed-strategy Nash equilibrium set is homeomorphic to a circle. More recently, \citet{biggar2026nash} investigate the computational difficulty of implementing Nash-convergent dynamics in nondegenerate finite games. In our setting, the nontrivial cohomology has a different origin. It arises from the \emph{projected first-order stationarity} conditions in smooth constrained games, where the normal-cone geometry of a box can produce cycles of FONE points even for affine fields associated with quadratic zero-sum payoffs. 

Beyond the topology of equilibrium sets, a complementary line of work studies the complex long-run behavior of learning dynamics. This includes Poincaré recurrence in regularized learning for zero-sum games \citep{piliouras2014optimization,mertikopoulos2018cycles}, nonconvergence of multiplicative weights \citep{bailey2018multiplicative}, and recurrence in replicator dynamics \citep{boone2019darwin}. Going further, \citet{andrade2023noregret} show that no-regret learning can simulate universal computation. These results illustrate that the existence of equilibria alone provides limited insight into the long-run behavior of learning dynamics.

Another line of work studies the informational limitations of equilibrium learning. In uncoupled dynamics, where each player observes only its own payoff function and the history of play, convergence to Nash equilibrium cannot be guaranteed across all games, even when the equilibrium is unique \citep{hart2003uncoupled}. Subsequent work investigates equilibrium learning under decentralized or payoff-based information, including the role of stochastic dynamics and memory \citep{hart2006stochastic,babichenko2012completely,jaggard2014self,toonsi2023higher}.

Complementing these dynamical and informational results, a separate line of work investigates the computational complexity of equilibrium problems. The introduction of PPAD by \citet{papadimitriou1992inefficient} provided a framework for studying fixed-point computation, and Nash equilibrium computation was subsequently shown to be PPAD-complete even in two-player finite games \citep{daskalakis2009complexity,chen2009settling}. For nonconvex--nonconcave min-max optimization, \citet{daskalakis2021complexity} proved PPAD-hardness under coupled constraints, while \citet{bernasconi2026complexity} extended these results to product constraints, including hypercubes. Recent work has further shown that quadratic objectives already suffice for such hardness. In particular, \citet{anagnostidescomplexity} proved PPAD-completeness for symmetric first-order equilibria in symmetric quadratic min-max problems, and later \citet{bernasconi2026quadratic} proved PPAD-hardness for computing approximate first-order stationary points of quadratic min-max objectives over hypercubes.

These works focus on whether an equilibrium can be computed efficiently. Our barrier, by contrast, concerns the possibility of stable global convergence, independently of computational resources. Even for a non-concave zero-sum game with quadratic payoffs and product hypercube constraints, no continuous learning dynamics that fix every FONE point can make the full FONE set a stable global attractor. Further related work is discussed in \Cref{sec:add_related}.

\section{Local Nash (FONE) and Projected Stationarity}
\label{sec:fone-prelim}

We consider a two-player differentiable game $G=(u_1,u_2)$ with compact convex feasible sets
$\mathcal{X}\subset \mathbb{R}^m$ and $\mathcal{Y}\subset \mathbb{R}^n$, and differentiable payoff
functions $u_1,u_2:\mathcal{X}\times\mathcal{Y}\to\mathbb{R}$. Both players are maximizers. We are particularly interested in the \emph{zero-sum} games where $u_2=-u_1$.

\begin{definition}[Local Nash equilibrium]
Fix any step size $\eta>0$. A profile $(x,y)\in\mathcal{X}\times\mathcal{Y}$ is a \emph{local Nash equilibrium} ({projected first-order Nash equilibrium}) if
\begin{equation}
\label{eq:fone-def}
    x=\Proj_{\mathcal{X}}\bigl(x+\eta\nabla_x u_1(x,y)\bigr),
    \qquad
    y=\Proj_{\mathcal{Y}}\bigl(y+\eta\nabla_y u_2(x,y)\bigr).
\end{equation}
The set of all such profiles is denoted by $\FONE(G)$, or equivalently
$\FONE(u_1,u_2)$.
\end{definition}

The choice of $\eta>0$ does not affect the exact solution set. Indeed, for any closed convex set
$K$, the projection identity $ z=\Proj_K(z+\eta g) $ is equivalent to the variational inequality
\[
\numberthis{eq:variational-ineq}
    \langle g,w-z\rangle \le 0
    \qquad \text{for all } w\in K.
\]
Equivalently, $g\in N_K(z)$, where
$N_K(z)=\{v:\langle v,w-z\rangle\le 0 \text{ for all } w\in K\}$
is the outward normal cone of $K$ at $z$. Since $N_K(z)$ is a cone, multiplying $g$ by any positive scalar leaves the exact first-order condition unchanged. The step size $\eta$ nevertheless matters when defining approximate projected residuals. For box constraints, the variational inequality admits the following coordinatewise characterization.

\begin{lemma}
\label{lem:box-rule}
Let $s\in[a,b]$, $g\in\mathbb{R}$, and $\eta>0$. Then
\[
    s=\Proj_{[a,b]}(s+\eta g)
\]
if and only if
\begin{equation}
\label{eq:box-rule}
    s=a \Rightarrow g\le 0,
    \qquad
    a<s<b \Rightarrow g=0,
    \qquad
    s=b \Rightarrow g\ge 0.
\end{equation}
\end{lemma}

At an interior coordinate, the corresponding component of the player's payoff gradient must vanish, while at a boundary it may point outward from the feasible interval. We apply \Cref{lem:box-rule} coordinatewise to characterize the exact $\FONE$ sets throughout the paper.

We also define the max projected residual
\begin{equation}
\label{eq:residual}
    \mathcal{R}^G_\eta(x,y)
    =
    \max\left\{
    \left\|x-\Proj_{\mathcal{X}}\bigl(x+\eta\nabla_x u_1(x,y)\bigr)\right\|_2,
    \left\|y-\Proj_{\mathcal{Y}}\bigl(y+\eta\nabla_y u_2(x,y)\bigr)\right\|_2
    \right\}.
\end{equation}
Then $
    \FONE(G)=\{(x,y):\mathcal{R}^G_\eta(x,y)=0\},
$
and the $\varepsilon$-approximate first-order Nash set is
$
    \AFONE_\varepsilon(G)
    =
    \{(x,y)\in\mathcal{X}\times\mathcal{Y}:
    \mathcal{R}^G_\eta(x,y)\le \varepsilon\}.
$
We write
$
\PStat(G)=\{z\in K:\ z=\Proj_K(z+\eta G(z))\}
$
for the projected stationary set of \(G\) on \(K\), equivalently \(G(z)\in N_K(z)\).

\section{General Homological Barrier to  Learning Dynamics}
\label{sec:master}

We begin with a general topological barrier for continuous deterministic learning dynamics, with applications beyond games, including optimization and fixed-point methods. Throughout, $\mathbb{F}$ denotes a fixed field, and $\check H^q(\cdot;\mathbb F)$ denotes \v{C}ech cohomology. We use \v{C}ech cohomology to accommodate potentially irregular attracting sets. For compact polyhedra, including the FONE sets constructed below, it agrees with singular cohomology, while its continuity property over neighborhoods plays a central role in our proof \citep{spanier1966algebraic,mardevsic1982shape}.

\begin{definition}\label{def:unif_attract}
Let $K$ be a compact metric space, let $A\subseteq K$ be compact, and let
$(T_t)_{t\ge 0}$ be a sequence of continuous maps $T_t:K\to K$. Define
\[
    \Psi_N = T_{N-1}\circ \cdots \circ T_0 .
\]
We say that $(T_t)_{t\ge 0}$ is \emph{pointwise-fixed and uniformly attracting
to $A$} if:
\begin{enumerate}[leftmargin=1.8em, itemsep=2pt, topsep=1pt, parsep=2pt, partopsep=2pt]
    \item $T_t(a)=a$ for every $a\in A$ and every $t\ge 0$; and
    \item for every open neighborhood $U\supseteq A$, there exists $N$ such that
    $\Psi_N(K)\subseteq U$.
\end{enumerate}
\end{definition}

Both conditions in \Cref{def:unif_attract} have natural interpretations for learning dynamics. The pointwise-fixed condition captures the usual stationarity requirement for an equilibrium solver, ensuring that once an iterate reaches a solution $a\in A$, all subsequent updates leave it unchanged. The second condition requires the dynamics to approach the target set $A$ uniformly over all initial conditions. The formulation includes autonomous iterations, corresponding to $T_t=T$ for all $t$, and continuous-time semiflows $(\Phi_t)_{t\ge 0}$ through a fixed time map $\Phi_T$. The topological barrier in the following theorem therefore applies to a broad class of learning algorithms, both in games and beyond, as discussed in \Cref{cor:no-lyapunov,cor:dynamics-variants,cor:finite-memory-gradient} and summarized in \Cref{tab:coverage}.

\begin{theorem}[Master Barrier Theorem]
\label{thm:master}
Let $K$ be compact and contractible, and let $A\subseteq K$ be compact. If
\[
    \check H^q(A;\mathbb{F})\neq 0
\]
for some $q\ge 1$, then there is no sequence of continuous maps
$T_t:K\to K$ that is pointwise-fixed and uniformly attracting to $A$.
\end{theorem}

\begin{proof}
Suppose, for contradiction, that such a sequence exists. Let $U\supseteq A$ be
an open neighborhood. By uniform attraction, there is an integer $N$ such that
$\Psi_N(K)\subseteq U$. Since each $T_t$ fixes $A$ pointwise, the composition
$\Psi_N$ also fixes $A$ pointwise.

Let $i:A\hookrightarrow K$ and $j:A\hookrightarrow U$ denote the inclusion maps.
Viewing $\Psi_N$ as a map from $K$ to $U$, we have $ \Psi_N\circ i = j $. By contravariant functoriality of \v{C}ech cohomology, $ j^* = i^*\circ \Psi_N^* $.
Since $K$ is contractible, $\check H^q(K;\mathbb{F})=0$ for every $q\ge 1$.
Therefore $j^*=0$. Thus, for every neighborhood $U$ of $A$, the canonical map $
    \check H^q(U;\mathbb{F}) \longrightarrow \check H^q(A;\mathbb{F}) $
is zero. Since \v{C}ech cohomology of $A$ is the direct limit of the cohomologies
of its neighborhoods, $
    \check H^q(A;\mathbb{F})
    \cong
    \varinjlim_{U\supseteq A} \check H^q(U;\mathbb{F}),
$
this forces $\check H^q(A;\mathbb{F})=0$, contradicting the hypothesis.
\end{proof}

The structure and proof of our barrier are closely related to classical shape-theoretic results connecting the shape of global attractors to that of the ambient phase space \citep{hastings1978shape,garay1991strong,kapitanski2000shape,milionis2023impossibility}. The following results show that, for dynamics that fix the target pointwise, Lyapunov stability together with pointwise attraction implies the uniform attraction required by \Cref{def:unif_attract}. Consequently, the barrier in \Cref{thm:master} applies to this broader class of dynamics.

\begin{table}[!t]
\centering
\small
\caption{Algorithmic interpretations of the master barrier in \Cref{thm:master}. 
}
\label{tab:coverage}

\begin{tabularx}{\textwidth}{@{}p{0.35\textwidth}p{0.13\textwidth}p{0.09\textwidth}X@{}}
\toprule
Algorithmic family & State space & Covered? & Reference\\
\midrule
Continuous-time gradient play 
& $K$ 
& yes 
& Semiflow (\Cref{cor:dynamics-variants}). \\

Projected GDA 
& $K$ 
& yes 
& (\Cref{cor:dynamics-variants}). \\

Extragradient / optimistic methods 
& $K$ or $K\times M$ 
& yes
& Finite-memory (\Cref{cor:finite-memory-gradient}). \\

Heavy-ball / momentum methods 
& $K\times M$ 
& yes 
& Contractible memory (\Cref{cor:finite-memory-gradient}). \\

Time-varying step sizes 
& $K$ 
& yes 
& Nonautonomous (\Cref{cor:dynamics-variants}). \\

Discontinuous or randomized methods 
& varies 
& no 
& (\Cref{remark:no}) \\

Algorithms moving along $\FONE(G)$ 
& varies 
& no 
& (\Cref{remark:no}) \\
\bottomrule
\end{tabularx}
\end{table}

\begin{lemma}
\label{lem:stability-uniform}
Let $T:K\to K$ be a continuous map on a compact metric space, and let
$A\subseteq K$ be compact and invariant. Suppose that $A$ is Lyapunov stable and
pointwise attracting; that is,
\[
    \operatorname{dist}(T^n z,A)\to 0
    \qquad \text{for every } z\in K .
\]
Then $A$ is uniformly attracting: for every open neighborhood $U\supseteq A$,
there exists $N$ such that $T^n(K)\subseteq U$ for all $n\ge N$.
\end{lemma}

\begin{proof}
Fix an open neighborhood $U\supseteq A$. By Lyapunov stability, there exists a
neighborhood $V\supseteq A$ such that every trajectory starting in $V$ remains in
$U$. For each $z\in K$, pointwise attraction gives an integer $n_z$ such that
$T^{n_z}z\in V$. By continuity of $T^{n_z}$, there is a neighborhood $W_z$ of
$z$ such that
$
    T^{n_z}(W_z)\subseteq V .
$
Compactness of $K$ gives a finite subcover
$W_{z_1},\ldots,W_{z_m}$ of $K$. Let $N=\max_i n_{z_i}$. If
$w\in W_{z_i}$ and $n\ge N$, then $T^{n_{z_i}}w\in V$, and Lyapunov stability
keeps all subsequent iterates in $U$. Hence $T^n(K)\subseteq U$ for all
$n\ge N$.
\end{proof}

\begin{corollary}[No strict Lyapunov-style certificate]
\label{cor:no-lyapunov}
Let $K$ be compact and contractible, and let $A\subseteq K$ be compact with
$\check H^q(A;\mathbb{F})\neq 0$ for some $q\ge 1$. There is no continuous map
$T:K\to K$ and continuous function $V:K\to \mathbb{R}_{\ge 0}$ satisfying:
\begin{enumerate}[leftmargin=1.8em, itemsep=2pt, topsep=1pt, parsep=2pt, partopsep=2pt]
    \item $T(a)=a$ for every $a\in A$;
    \item $V^{-1}(0)=A$; and
    \item $V(Tz)<V(z)$ for every $z\notin A$.
\end{enumerate}
The same conclusion holds for a continuous-time semiflow whenever its time-one
map satisfies the strict decrease condition. Consequently, no global strict
Lyapunov argument can certify convergence to such a pointwise-fixed target.
\end{corollary}

\begin{proof}
We show that such a Lyapunov certificate would imply uniform attraction, thereby
contradicting \Cref{thm:master}. Let $U\supseteq A$ be open. Since
$V^{-1}(0)=A$ and $K\setminus U$ is compact and disjoint from $A$, $
    a_U := \min_{z\in K\setminus U} V(z) > 0 .$
On the compact set $\{z\in K: V(z)\ge a_U\}$, the continuous function
$z\mapsto V(z)-V(Tz)$ is strictly positive. Hence it has a positive minimum,
say $\delta_U>0$. Therefore, as long as $V(T^n z)\ge a_U$, the Lyapunov value
decreases by at least $\delta_U$ at the next step. Since $V$ is bounded on the
compact set $K$, every orbit enters $\{V<a_U\}$ after a uniformly bounded
number of steps. Because $\{V<a_U\}\subseteq U$, the target $A$ is pointwise-fixed
and uniformly attracting, contradicting \Cref{thm:master}.
\end{proof}

\begin{corollary}[Obstructed dynamics]
\label{cor:dynamics-variants}
Let $K$ and $A$ be as in \Cref{thm:master}. Then $A$ cannot be:
\begin{enumerate}[leftmargin=1.8em]
    \item the pointwise-fixed uniformly attracting target of a continuous
    semiflow on $K$;
    \item the pointwise-fixed uniformly attracting target of an autonomous
    continuous map $T:K\to K$;
    \item the pointwise-fixed uniformly attracting target of a time-varying
    deterministic algorithm on $K$;
    \item the pointwise-fixed uniformly attracting target $B\subseteq K\times M$
    of a finite-memory algorithm, whenever $M$ is compact and contractible and
    $\check H^q(B;\mathbb{F})\neq 0$ for some $q\ge 1$.
\end{enumerate}
In particular, if the intended memory target is $B=A\times\{m_\star\}$ with $m_\star$ as in \Cref{cor:finite-memory-gradient}, then $
    \check H^q(B;\mathbb{F})
    \cong
    \check H^q(A;\mathbb{F})$,
so adding compact contractible memory does not remove the barrier.
\end{corollary}

\begin{corollary}[Finite-memory gradient methods]
\label{cor:finite-memory-gradient}
Let $G$ be a game with compact contractible strategy space $K$ and FONE target
$A=\FONE(G)$. Suppose that $
    \check H^q(A;\mathbb{F})\neq 0
$
for some $q\ge 1$. Consider any deterministic finite-memory method with compact
contractible memory $M$ and continuous update rule
\[
    (z_{t+1},m_{t+1}) = T_G(z_t,m_t),
    \qquad
    (z_t,m_t)\in K\times M .
\]
Let $m_\star\in M$ be a distinguished consistent-memory state. If
\[
    T_G(a,m_\star)=(a,m_\star)
    \qquad \text{for every } a\in A,
\]
then the target $A\times\{m_\star\}$ cannot be uniformly globally attracting in
$K\times M$.

In particular, projected GDA with finitely many auxiliary variables,
extragradient and optimistic variants represented as finite-memory methods, and
momentum methods cannot admit a universal Lyapunov-style convergence proof that
both fixes every FONE and uniformly attracts to $\FONE(G)\times\{m_\star\}$ on
such games.
\end{corollary}

\begin{remark}
\label{remark:no}
The barrier applies to continuous deterministic dynamics on compact contractible state spaces that fix every point of the target and attract the full target uniformly, or satisfy the corresponding Lyapunov stability and pointwise-attraction conditions. Randomized or discontinuous algorithms, methods that converge to a selected FONE point, and dynamics that move along the FONE set fall outside these assumptions. For finite-memory methods with full state $(z,m)\in K\times M$, where $M$ is compact and contractible, the barrier applies when the stationary target in the extended space has the same topology as $\FONE(G)$, as in $\FONE(G)\times\{0\}$. In contrast, a method may use memory to construct an extended attractor $B$ whose projection onto strategy space is $\FONE(G)$ but whose total-space topology is contractible. Such a construction falls outside the scope of the barrier.
\end{remark}

The proof of this result and further discussion appear in \Cref{app:barrier-details,app:memory-app}. In view of \Cref{thm:master,cor:finite-memory-gradient}, our focus and proof strategy in the following sections is to construct games whose FONE sets have nontrivial positive-degree \v{C}ech cohomology, namely $\check H^q(\FONE;\mathbb{F})\neq 0$ for some degree $q\ge 1$.

\section{Homological Barrier for Local (First-order) Nash}

In this section, we instantiate the barrier from \Cref{sec:master} by constructing games whose FONE sets have nontrivial topology. Such sets cannot be stable global attractors for continuous pointwise-stationary learning dynamics.

\begin{tcolorbox}[
  colback=maroon!8!white,
  colframe=heavymaroon,
  boxrule=1pt,
  arc=2pt,
  left=2mm,right=2mm,top=1mm,bottom=1mm
]
A nonconcave-concave quadratic zero-sum game on a box can have a FONE set with nontrivial first homology, as in the presence of a loop. Consequently, its exact FONE set cannot be a stable global attractor for any continuous pointwise-stationary learning rule.
\end{tcolorbox}

We first present a degree-$6$ nonconcave-nonconcave scalar game as a warm-up to illustrate the underlying geometry. We then realize the same barrier in quadratic zero-sum games, showing that it can arise even at the lowest nontrivial polynomial degree.

\subsection{Warm-up: A Minimal Example with Topologically Nontrivial FONE set}
\label{sec:warmup}

We begin with a simple example that illustrates the underlying geometry. To apply the topological barrier in \Cref{thm:master}, we seek to construct a game whose FONE set contains a circle. However, a closed curve of interior first-order stationary points necessarily forces additional stationary points in the region it encloses. The following elementary lemma explains why such a circle cannot form the entire FONE set.

\begin{lemma}
\label{lem:forced-point}
Let $f\in C^1(D^2)$. If $\nabla f=0$ on $\partial D^2$, then either $f$ is constant on $D^2$ or $f$ has a critical point in the interior.
\end{lemma}

\begin{proof}
Since the tangential derivative vanishes on $\partial D^2$, the function is constant along the boundary. If $f$ is not constant on $D^2$, then either its maximum or its minimum differs from the boundary value. By compactness, such an extremum is attained in the interior, and Fermat's rule implies the existence of an interior critical point.
\end{proof}

In particular, this result shows that any construction that enforces stationarity along a closed curve must also include at least one FONE point in the interior of that curve. We now design such a game.

Both players choose scalar strategies in $[-2,2]$. Define $
    h(x,y)=x^2+y^2-1,
    k(x,y)=1+y^2-x^2,
$
and consider the zero-sum game
\begin{equation}
\label{eq:warmup-payoff}
    u_1(x,y)= h(x,y)^2 k(x,y),
    \qquad
    u_2(x,y)=-u_1(x,y).
\end{equation}
The factor $h^2$ enforces first-order stationarity on the unit circle $\{x^2+y^2=1\}$, while the factor $k$ eliminates all additional stationary points except for a single unavoidable interior point.

\begin{figure}[t]
\centering

\begin{subfigure}[t]{0.48\textwidth}
\centering
\begin{tikzpicture}[scale=.82]
    \draw[thin] (-2,-2) rectangle (2,2);
    \draw[->,gray] (-2.35,0)--(2.35,0) node[right] {$x$};
    \draw[->,gray] (0,-2.35)--(0,2.35) node[above] {$y$};

    \draw (-2,0.06)--(-2,-0.06) node[below left=3pt and 1pt] {\scriptsize $-2$};
    \draw (-1,0.06)--(-1,-0.06) node[below left=3pt and 1pt] {\scriptsize $-1$};
    \draw (1,0.06)--(1,-0.06) node[below right=3pt and 1pt] {\scriptsize $1$};
    \draw (2,0.06)--(2,-0.06) node[below right=3pt and 1pt] {\scriptsize $2$};

    \draw (0.06,2)--(-0.06,2) node[above left=1pt and 3pt] {\scriptsize $2$};
    \draw (0.06,1)--(-0.06,1) node[above left=1pt and 3pt] {\scriptsize $1$};
    \draw (0.06,-1)--(-0.06,-1) node[below left=1pt and 3pt] {\scriptsize $-1$};
    \draw (0.06,-2)--(-0.06,-2) node[below left=1pt and 3pt] {\scriptsize $-2$};

    \draw[very thick,blue] (0,0) circle (1);
    \fill[red] (0,0) circle (2.1pt);
    \node[blue,above right] at (0.03,0.99) {\small $x^2+y^2=1$};
    \node[red,below right] at (-0.1,+0.1) {\small $(0,0)$};
\end{tikzpicture}
\caption{The exact FONE set of the game~\eqref{eq:warmup-payoff}.}
\label{fig:warmup}
\end{subfigure}
\hfill
\begin{subfigure}[t]{0.5\textwidth}
\centering
\begin{tikzpicture}[scale=.72, line cap=round, line join=round]
  \coordinate (000) at (0,0);
  \coordinate (100) at (3,0);
  \coordinate (110) at (3,3);
  \coordinate (010) at (0,3);

  \coordinate (001) at (0.9,0.8);
  \coordinate (101) at (3.9,0.8);
  \coordinate (111) at (3.9,3.8);
  \coordinate (011) at (0.9,3.8);

  \draw[gray!55, thin] (001)--(101)--(111)--(011)--cycle;
  \draw[gray!55, thin] (000)--(001);
  \draw[gray!55, thin] (100)--(101);
  \draw[gray!55, thin] (110)--(111);
  \draw[gray!55, thin] (010)--(011);

  \draw[gray!55, thin] (000)--(100)--(110)--(010)--cycle;

  \draw[very thick,blue]
    (000)--(100)--(110)--(111)--(011)--(001)--cycle;

  \fill[red] (1.95,1.9) circle (2pt);
  \node[right] at (1.95,1.9) {\scriptsize $p$};

  \node[below left] at (000) {\scriptsize $(0,0,0)$};
  \node[below] at (100) {\scriptsize $(1,0,0)$};
  \node[right] at (110) {\scriptsize $(1,1,0)$};
  \node[right] at (111) {\scriptsize $(1,1,1)$};
  \node[above] at (011) {\scriptsize $(0,1,1)$};
  \node[left] at (001) {\scriptsize $(0,0,1)$};
\end{tikzpicture}
\caption{The quadratic projected-gradient gadget of \eqref{eq:phi}.}
\label{fig:cube}
\end{subfigure}

\caption{Two FONE geometries, the circles carry the topological barrier. Left: the unit circle together with the isolated origin. Right: the six-edge cycle \(\calC\subset[0,1]^3\) with isolated interior zero \(p\).}
\label{fig:fone-geometries}
\end{figure}
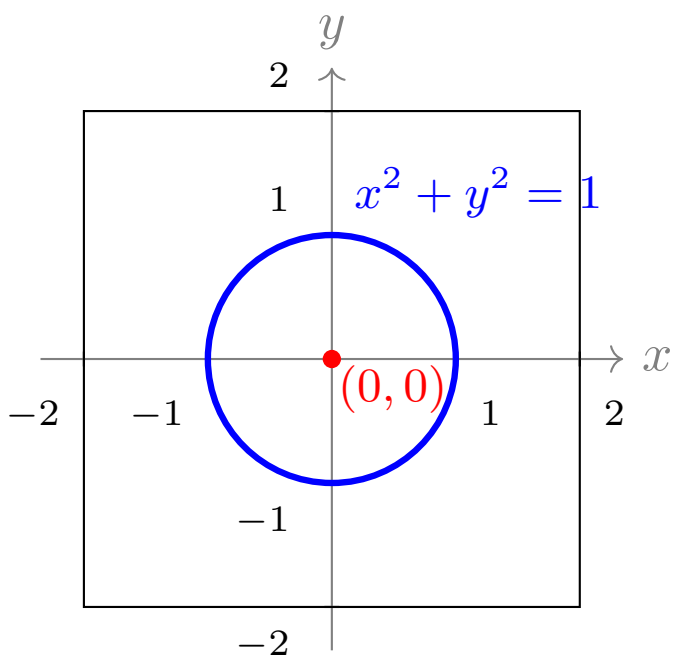
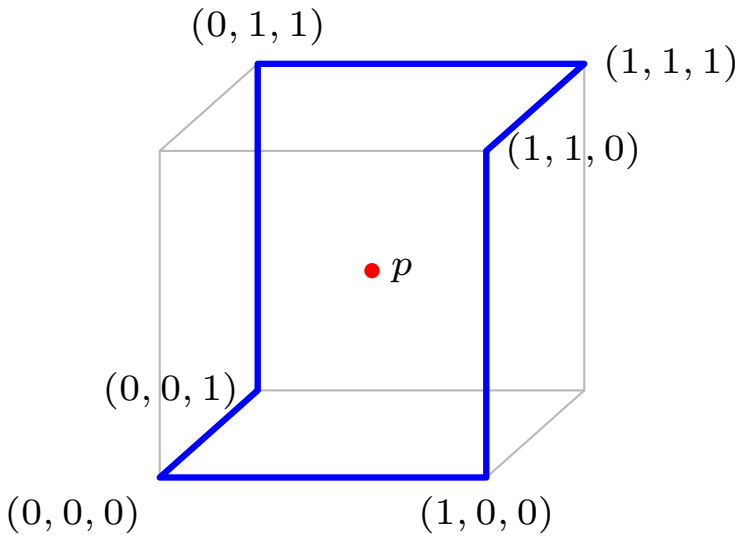

\begin{theorem}
\label{thm:warmup}
For the game \eqref{eq:warmup-payoff},
\[
    \FONE(u_1,u_2)
    =
    \{(x,y):x^2+y^2=1\}\ \cup\ \{(0,0)\}.
\]
In particular, $H_1(\FONE;\mathbb{Z})\cong \mathbb{Z}$.
\end{theorem}

We defer the proof of this result to \Cref{app:missing_forced}. As a consequence of \Cref{thm:warmup}, even this scalar zero-sum example yields a FONE set with nontrivial topology as depicted in \Cref{fig:warmup}. Since $\check H^1(\FONE;\mathbb{F})\neq 0$, \Cref{thm:master} (with $q=1$) rules out any continuous pointwise-stationary dynamics that globally attracts to $\FONE$.

\subsection{A Quadratic Zero-sum Game with Homological barrier}

We now show that the topological barrier identified in \Cref{thm:master}
already arises in quadratic zero-sum games. The key idea is to first construct
a projected-gradient gadget whose stationary set has nontrivial topology, and
then embed this structure into a two-player game. This yields a quadratic
zero-sum game whose FONE set inherits the same topological barrier,
thereby ruling out continuous pointwise-stationary dynamics that uniformly
attract to it.

\subsubsection{A Quadratic Projected-Gradient Gadget}
\label{sec:gadget}
We now construct a quadratic gadget whose projected stationary set has nontrivial topology. Unlike the warm-up construction, where the stationary curve lies in the interior of the feasible domain and necessarily forces an additional critical point (\Cref{lem:forced-point}), here the cycle is realized entirely on the boundary of the cube.

A natural attempt to exploit nonconcavity is to consider the gradient field induced by maximizing the convex function $\psi(x)=\|x\|^2$. However, as illustrated in \Cref{fig:pstat-positive-norm} in \Cref{app:details}, this does not produce the required nontrivial topology when restricted to the hypercube, since $H_1(\PStat(\nabla \psi);\Z)=0$. Therefore, a different construction is required. Let
\begin{equation}
\label{eq:phi}
    \phi(z_1,z_2,z_3)=z_1z_2 - z_1z_3 + z_2z_3 - z_2, \quad \textrm{and} \quad G(z)=\nabla\phi(z)=
    \begin{pmatrix}
      z_2 - z_3\\
      z_1 + z_3 - 1\\
      z_2 - z_1
    \end{pmatrix}.
\end{equation}
Define the six-edge boundary cycle
\begin{align}
\label{eq:cycle}
  \calC={}&
  \{(t,0,0):0\le t\le1\}
  \cup\{(1,t,0):0\le t\le1\}
  \cup\{(1,1,t):0\le t\le1\} \nonumber\\
  &\cup\{(t,1,1):0\le t\le1\}
  \cup\{(0,t,1):0\le t\le1\}
  \cup\{(0,0,t):0\le t\le1\},
\end{align}
and let $p=(1/2,1/2,1/2)$. \Cref{fig:cube} illustrates the projection of the stationary set of the affine field $\nabla \phi(z)$ onto the hypercube.

\begin{lemma}
\label{lem:gadget}
For the affine field $G=\nabla\phi$ on $[0,1]^3$, the projected stationary set is exactly
\[
    \PStat(G)=\calC\cup\{p\}.
\]
Consequently $H_1(\PStat(G);\Z)\cong\Z$.
\end{lemma}

\begin{proof}
Each edge of $\calC$ satisfies the box sign conditions. For instance, on $E_1=\{(t,0,0):0\le t\le1\}$, the first coordinate is interior and $G_1=0$, while the second and third coordinates are at lower endpoints with $G_2=t-1\le0$ and $G_3=-t\le0$. The remaining edges follow by identical checks. The point $p$ lies in the interior and satisfies $G(p)=0$, hence $
\calC\cup\{p\}\subseteq\PStat(G)$.

Let $z=(a,b,c)\in\PStat(G)$: i) If $0<b<1$, then coordinate $2$ is interior, so $G_2=0$, i.e., $a+c=1$. If $a,c\in(0,1)$, then $G_1=G_3=0$, implying $a=b=c=1/2$, i.e., $z=p$. If $a=0$, then $c=1$, giving $(0,b,1)\in E_5$; if $a=1$, then $c=0$, giving $(1,b,0)\in E_2$, ii) if $b=0$, the lower-endpoint condition gives $a+c-1\le0$. Interior conditions on $a$ or $c$ force $c=0$ or $a=0$, yielding $E_1\cup E_6$. iii) if $b=1$, the upper-endpoint condition gives $a+c-1\ge0$. Interior conditions force $c=1$ or $a=1$, yielding $E_3\cup E_4$. These cases exhaust $[0,1]^3$, so $
\PStat(G)=\calC\cup\{p\}$.

Thus, the set $\calC$ is a polygonal circle and $p$ is isolated, so $H_1\cong\Z$.
\end{proof}

\begin{remark}
This gadget shows that topologically nontrivial projected stationary sets arise already in smooth quadratic optimization over a hypercube. In particular, for the quadratic function $\phi$ on $[0,1]^3$, the projected stationary set is $\mathcal C\cup\{p\}$ and has nontrivial first homology. Therefore, by \Cref{thm:master}, no continuous pointwise-stationary projected dynamics can make the full projected stationary set a uniformly attracting global target. For example, projected gradient descent (PGD) is pointwise-stationary on $\PStat(\nabla\phi)$, but $\PStat(\nabla\phi)$ is not a stable global attractor. Concretely, it is easy to observe that for any neighborhood $U\supseteq \PStat(\nabla\phi)$, there exist initial points arbitrarily close to $p$ whose iterates leave $U$ under PGD. Thus, although PGD fixes every point in $\PStat(\nabla\phi)$, it fails to uniformly attract to the full stationary set.
\end{remark}

\subsubsection{Diagonal Forcing and the Quadratic Zero-sum Game}
\label{sec:quadratic}

We now lift the projected-gradient gadget to a two-player zero-sum game by diagonal forcing. Let $x,y\in[0,1]^3$, and define
\begin{equation}
\label{eq:main-game}
    u_1(x,y)=\phi(x)+\tfrac12\|x-y\|_2^2,
    \qquad
    u_2(x,y)=-u_1(x,y),
\end{equation}
where $\phi$ is given in \eqref{eq:phi}. The payoff is quadratic and zero-sum, and in contrast to the game developed in the warm-up~\eqref{eq:warmup-payoff}, it is even nonconcave-concave. The diagonal forcing induced by $\tfrac12\|x-y\|_2^2$ transfers the geometry of stationarity from the quadratic gadget in \Cref{sec:gadget} to the FONE set of the quadratic zero-sum game. This is formalized in the following result.

\begin{theorem}[FONE set of our quadratic zero-sum game]
\label{thm:main}
For the quadratic zero-sum game \eqref{eq:main-game},
\[
    \FONE(u_1,u_2)=\{(z,z):z\in\calC\}\cup\{(p,p)\}.
\]
Consequently $H_1(\FONE;\Z)\cong\Z$ and, by the universal coefficient theorem, $H^1(\FONE;\bbF)\cong\bbF$ for every field $\bbF$.
\end{theorem}

Therefore, we can invoke the master barrier theorem from \Cref{thm:master} for this quadratic game.

\begin{corollary}[Impossibility for quadratic games]
\label{cor:main-impossibility}
For the quadratic zero-sum game \eqref{eq:main-game}, no pointwise-fixed, continuous, uniformly attracting dynamics of the type covered by \Cref{thm:master,cor:dynamics-variants} can globally learn the exact FONE set. In particular, no Lyapunov-style argument satisfying \Cref{cor:no-lyapunov} can establish convergence to all FONE points.
\end{corollary}

We conclude this section by showing that degree-two zero-sum games form the minimal nonconcave class that exhibits such homological barriers for learning dynamics to find FONE.

\begin{proposition}[Degree two is minimal]
\label{prop:minimal}
For any two-player zero-sum game on a box whose payoff has total degree at most one, the FONE set is a product of faces of the box. In particular, it is contractible and has trivial first homology.
\end{proposition}

Full proofs of \Cref{thm:main,cor:main-impossibility,prop:minimal} are deferred to \Cref{app:missing_diagonal} for brevity.

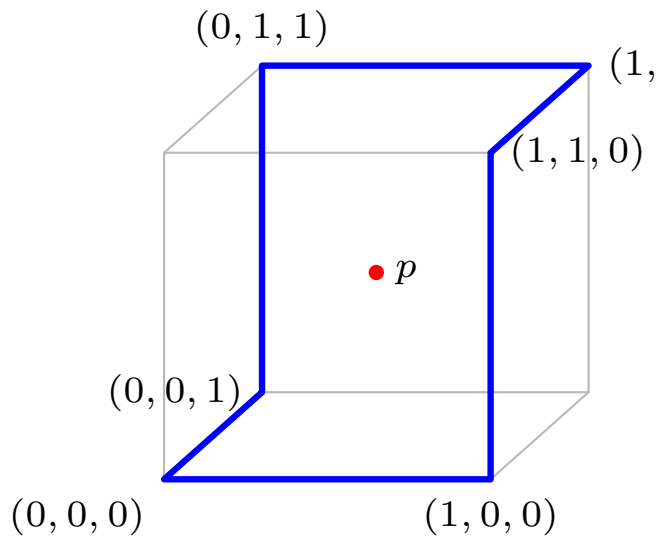
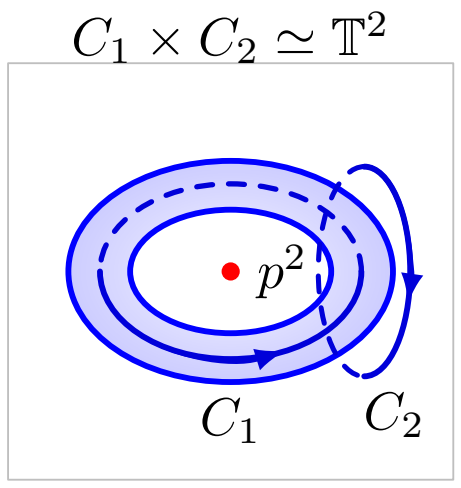
\begin{figure}[t]
\centering

\begin{subfigure}[t]{0.47\textwidth}
\centering
\begin{tikzpicture}[scale=.72, line cap=round, line join=round]
  \coordinate (000) at (0,0);
  \coordinate (100) at (3,0);
  \coordinate (110) at (3,3);
  \coordinate (010) at (0,3);

  \coordinate (001) at (0.9,0.8);
  \coordinate (101) at (3.9,0.8);
  \coordinate (111) at (3.9,3.8);
  \coordinate (011) at (0.9,3.8);

  \draw[gray!55, thin] (001)--(101)--(111)--(011)--cycle;
  \draw[gray!55, thin] (000)--(001);
  \draw[gray!55, thin] (100)--(101);
  \draw[gray!55, thin] (110)--(111);
  \draw[gray!55, thin] (010)--(011);

  \draw[gray!55, thin] (000)--(100)--(110)--(010)--cycle;

  \draw[very thick,blue]
    (000)--(100)--(110)--(111)--(011)--(001)--cycle;

  \fill[red] (1.95,1.9) circle (2pt);
  \node[right] at (1.95,1.9) {\scriptsize $p$};

  \node[below left] at (000) {\scriptsize $(0,0,0)$};
  \node[below] at (100) {\scriptsize $(1,0,0)$};
  \node[right] at (110) {\scriptsize $(1,1,0)$};
  \node[right] at (111) {\scriptsize $(1,1,1)$};
  \node[above] at (011) {\scriptsize $(0,1,1)$};
  \node[left] at (001) {\scriptsize $(0,0,1)$};
\end{tikzpicture}
\caption{One block: the quadratic gadget gives the six-edge cycle $C\subset[0,1]^3$ together with an isolated point $p$.}
\end{subfigure}
\hfill
\begin{subfigure}[t]{0.47\textwidth}
\centering
\begin{tikzpicture}[scale=0.95, line cap=round, line join=round]
    \useasboundingbox (-0.45,-0.35) rectangle (3.55,3.25);

    \draw[black!25, line width=0.35pt]
        (0.00,0.00) rectangle (3.10,2.90);

    \coordinate (T) at (1.55,1.45);

    \shade[even odd rule, inner color=blue!4, outer color=blue!20]
        (T) ellipse [x radius=1.13cm, y radius=0.77cm]
        (T) ellipse [x radius=0.70cm, y radius=0.43cm];

    \draw[blue, line width=1.10pt]
        (T) ellipse [x radius=1.13cm, y radius=0.77cm];
    \draw[blue, line width=1.10pt]
        (T) ellipse [x radius=0.70cm, y radius=0.43cm];

    \draw[blue!85!black, line width=1.00pt, dashed]
        ($(T)+(0.91,0)$)
        arc[start angle=0,end angle=180,x radius=0.91cm,y radius=0.61cm];
    
    \draw[blue!85!black, line width=1.10pt]
        ($(T)+(-0.91,0)$)
        arc[start angle=180,end angle=360,x radius=0.91cm,y radius=0.61cm];
    
    \draw[blue!85!black, line width=0.8pt, -{Latex[length=2mm]}]
        ($(T)+(-0.70,-0.39)$)
        arc[start angle=220,end angle=295,x radius=0.91cm,y radius=0.61cm];
    
    \node[black] at ($(T)+(0,-1.04)$) {$C_1$};

    \coordinate (M) at ($(T)+(0.93,0)$);

    \draw[blue!85!black, line width=0.95pt, dashed]
        ($(M)+(0,0.73)$)
        arc[start angle=90,end angle=270,x radius=0.32cm,y radius=0.73cm];

    \draw[blue!85!black, line width=1.10pt]
        ($(M)+(0,0.73)$)
        arc[start angle=90,end angle=-90,x radius=0.32cm,y radius=0.73cm];

    \draw[blue!85!black, line width=0.95pt, -{Latex[length=2mm]}]
        ($(M)+(0,0.73)$)
        arc[start angle=90,end angle=-18,x radius=0.32cm,y radius=0.73cm];

    \node[black] at ($(T)+(1.14,-1)$) {$C_2$};

    \node[black] at (1.55,3.08)
        {$C_1 \times C_2 \simeq \mathbb{T}^2$};

    \fill[red] (T) circle (1.8pt);
    \node[fill=white, inner sep=0.8pt, right=4pt] at (T) {$p^2$};

\end{tikzpicture}
\caption{Two independent copies: the component $C_1\times C_2$ is shown as a torus. 
}
\end{subfigure}

\caption{Product construction. One block has a cycle component $C$ and an isolated point $p$. With two independent copies, the topology-carrying component becomes $C_1\times C_2\simeq\mathbb{T}^2$, shown schematically as a torus with a visible hole. The point $p^2=(p,p)$ remains as a separate isolated component. The mixed circle components $C_1\times\{p\}$ and $\{p\}\times C_2$ are omitted from the schematic.}
\label{fig:product-construction}
\end{figure}

\subsection{Arbitrarily High-dimensional Homological Barriers}
\label{sec:block-product}

The single-cycle construction extends to higher-dimensional topology by combining independent copies of the quadratic gadget. The resulting zero-sum games have quadratic payoffs, while their FONE sets contain higher-dimensional torus components and have nontrivial homology in arbitrarily high degrees.

For $r\ge 1$, write $z=(z^{(1)},\ldots,z^{(r)})\in[0,1]^{3r}$, with each block $z^{(j)}\in[0,1]^3$. Define the block-separable quadratic potential
\[
\Phi_r(z)=\sum_{j=1}^r \phi(z^{(j)}).
\]
Its gradient acts independently on each block, giving the direct product of $r$ copies of $\nabla\phi$.

\begin{theorem}[Quadratic games with arbitrary homological degree]
\label{thm:block-product}
For every $r\ge1$, the quadratic zero-sum game
\[
    u_1^{(r)}(x,y)=\Phi_r(x)+\tfrac12\|x-y\|_2^2,
    \qquad
    u_2^{(r)}=-u_1^{(r)},
\]
has exact FONE set
\begin{equation}
\label{eq:block-fone}
    \FONE(u_1^{(r)},u_2^{(r)})=
    \{(z,z):z\in(\calC\cup\{p\})^r\}.
\end{equation}
Consequently,
\[
    H_r\bigl(\FONE(u_1^{(r)},u_2^{(r)});\Z\bigr)\ne0,
    \qquad
    \check H^r\bigl(\FONE(u_1^{(r)},u_2^{(r)});\bbF\bigr)\ne0.
\]
In particular, quadratic zero-sum games realize homological barriers in arbitrarily high degree.
\end{theorem}

\begin{proof}
Since $\Phi_r$ is block-separable, projected stationarity for $\nabla\Phi_r$ decouples across blocks. By \Cref{lem:gadget}, each block has projected stationary set $\calC\cup\{p\}$, so $ \PStat(\nabla\Phi_r)=(\calC\cup\{p\})^r.$ Applying \Cref{lem:diagonal} in dimension $3r$ yields \eqref{eq:block-fone}. The subset $\calC^r\subset(\calC\cup\{p\})^r$ is homeomorphic to the $r$-torus $\mathbb{T}^r$, and $H_r(\mathbb{T}^r;\Z)\cong\Z$. Since $\calC^r$ is a connected component of the FONE set, it follows that $H_r(\FONE;\Z)\ne0$. For compact polyhedra, singular and \v{C}ech cohomology agree, and the universal coefficient theorem over a field yields $\check H^r\ne0$.
\end{proof}

The theorem extends the barrier in \Cref{cor:main-impossibility} to arbitrarily high cohomological degrees, with the obstruction arising in degree $r$. A schematic illustration of this construction for the degree $r=2$ is given in \Cref{fig:product-construction}.

\subsection{Robust Approximate barrier}
\label{sec:robust}

We now show that the topological barrier extends to approximate-FONE targets and small perturbations of the game. For sufficiently small residual tolerances, the approximate-FONE set contains the core cycle and remains within a regular neighborhood where the cycle is cohomologically essential. We then prove that the barrier also holds under small $C^1$ perturbations, covering an open, positive-volume family of quadratic games.

Consider the quadratic game $G_\star$ defined in \eqref{eq:main-game}, with projected residual $\calR_\eta^\star$. Let
\[
\Delta\calC=\{(z,z):z\in\calC\}
\]
denote the diagonal copy of the cycle. Choose a tubular neighborhood $T$ of $\Delta\calC$ and a small closed ball $B$ around $(p,p)$, disjoint from $T$, and set $O=\operatorname{int}(T\cup B)$. Define the residual gap
\[
m_O\coloneqq\min_{w\in K\setminus O}\calR_\eta^\star(w)>0,
\qquad K=[0,1]^3\times[0,1]^3.
\]
The strict positivity follows from the continuity of $\calR_\eta^\star$, the compactness of $K\setminus O$, and the fact that its zero set is precisely $\FONE(G_\star)\subset O$. We use this gap to show that sufficiently small residual sublevel sets remain confined to $O$, where the diagonal cycle continues to provide a topological barrier.

\begin{theorem}[Approximate FONE and one-cycle]
\label{thm:approx}
For every $0<\eps<m_O$,
\[
    \Delta\calC\subseteq \AFONE_\eps(G_\star)\subseteq O.
\]
Moreover, $\check H^1(\AFONE_\eps(G_\star);\bbF)\ne0$. Consequently, $\AFONE_\eps(G_\star)$ cannot be a pointwise-fixed uniformly attracting target on the contractible strategy cube.
\end{theorem}

In turn, we show that our homological barrier extends to perturbed versions of the games we construct; consequently, it applies to a positive-volume set of quadratic games.

\begin{theorem}
\label{thm:c1}
Let $\widetilde G$ be a game whose own-gradients are $\delta$-close to those of $G_\star$, and let $\tau=\eta\delta$. If $0<\tau<m_O/4$ and
$ \tau\le\eps<m_O-\tau $, then $
\Delta\calC\subseteq\AFONE_\eps(\widetilde G)\subseteq O,
$ and $\check H^1(\AFONE_\eps(\widetilde G);\bbF)\ne0$. Hence these approximate targets cannot be pointwise-fixed uniformly attracting targets.
\end{theorem}

\begin{corollary}
\label{cor:open-coefficient}
Fix any finite-dimensional parameterization of quadratic zero-sum games containing $G_\star$. There exists an open neighborhood $\calU$ of $G_\star$ such that for every $\widetilde G\in\calU$, there is a nonempty interval $I_{\widetilde G}$ with
\[
    \check H^1(\AFONE_\eps(\widetilde G);\bbF)\ne0
    \quad \text{for all } \eps\in I_{\widetilde G}.
\]
Consequently, sufficiently small approximate-FONE targets fail to be uniformly attracting on an open, positive-volume set of quadratic games.
\end{corollary}

These results establish that the barrier is robust to both approximation and small perturbations. In particular, it is not a pathological artifact of a single construction, but instead characterizes an open set of quadratic zero-sum games of positive measure.

\section{Conclusion}

We showed that stable global learning of first-order Nash equilibria faces a topological barrier even in quadratic zero-sum games. Starting from an affine projected-gradient field on a cube with a stationary six-edge cycle, we construct a quadratic zero-sum game whose FONE set consists of this cycle and an isolated point. The nontrivial topology of this set rules out continuous pointwise-stationary dynamics that make the entire FONE set a stable global attractor on a compact contractible state space. We further show that the barrier extends to sufficiently small approximate-FONE targets and nearby games. More broadly, our results demonstrate how projection and boundary normal cones can give rise to homologically nontrivial stationary sets even for affine fields. This geometric mechanism may also have implications for constrained nonconvex optimization, variational inequalities, complementarity problems, dynamical systems, and multi-agent control.

\section*{Acknowledgement}

A.S. and P.J. were partially supported by the ONR grant N00014-24-1-2470. Part of this research was performed while A.S. was visiting the Institute for Mathematical and Statistical Innovation (IMSI), which is supported by the National Science Foundation (Grant No. DMS-2425650). G.F. was supported in part by the National Science Foundation awards CCF-2443068 and IIS-2552046, the Office of Naval Research grant N000142512296, and an AI2050 Early Career Fellowship.

\bibliographystyle{unsrtnat}
\bibliography{refs} 

\newpage
\appendix

\section{Nonconvergence and Instability for Learning in Games}
\label{sec:add_related}

A substantial body of work in game dynamics documents failures of equilibrium convergence, even when the equilibrium notion is well defined. For FTRL-type no-regret learning in finite games, \citet{vlatakis2020no} show that stable attracting Nash equilibria must be strict, and hence pure, ruling out stable attraction to mixed Nash equilibria. Related limitations arise in min--max optimization. \citet{hsieh2021limits} show that broad classes of stochastic-approximation schemes can converge to internally chain-transitive sets containing no critical points, while \citet{letcher2021impossibility} construct analytic zero-sum multi-loss examples in which reasonable learning algorithms must either diverge or cycle.  These works point to
a common theme that natural learning rules can converge to the wrong objects, or
fail to converge at all.

A related line of work investigates the recurrent and chaotic behavior of familiar learning dynamics across different game-theoretic settings. In congestion games, multiplicative weights can produce cycles and chaotic trajectories \citep{palaiopanos2017multiplicative}. For differentiable games, decompositions into potential-like and Hamiltonian-like components help explain conservation laws and recurrent motion \citep{balduzzi2018mechanics,nagarajan2020chaos}. Similarly, no-regret and regularized learning in adversarial games can lead to sustained cycling and Poincar\'e recurrence \citep{piliouras2014optimization,mertikopoulos2018cycles}. Such behavior also arises in applied multi-agent settings, including performative prediction, autobidding auctions, and linear-quadratic games, where simple learning dynamics can transition from convergence to periodicity or chaos \citep{chotibut2020route,mazumdar2020policy,bielawski2021follow,piliouras2023multi,paes2024complex}. Together, these findings motivate studying the long-run behavior of learning dynamics, which can reveal fundamental properties of games that are not apparent from their equilibrium sets alone.

The preceding results concern the behavior of particular learning rules. Our barrier instead arises from the topology of the target set and applies even to fully coupled continuous dynamics with complete knowledge of the game. In our quadratic zero-sum construction, the FONE set
\[
\FONE(G_\star)=\{(z,z):z\in\mathcal C\}\cup\{(p,p)\}
\]
has nontrivial first homology, while the ambient strategy space is contractible. Consequently, no continuous dynamics can both fix every FONE point and uniformly attract to the entire FONE set. This limitation is independent of the update rule and does not rely on any particular cycling or divergent behavior.

\section{Preliminaries}
\label{app:prelim}

In this section, we briefly review the preliminaries used throughout the paper. All homology and cohomology groups are taken with coefficients in a fixed field $\bbF$, unless integer coefficients are explicitly specified.

\subsection{Projection, Normal Cones, and Box Stationarity}
\label{app:projection}

We use standard facts from convex analysis and variational inequalities, and refer to \citet{rockafellar1998variational,facchinei2003finite} for standard references. Let $K\subseteq\R^d$ be nonempty, closed, and convex. The Euclidean projection
\[
    \Proj_K(z)=\arg\min_{w\in K}\|z-w\|_2
\]
exists and is unique.

\begin{lemma}
\label{lem:projection-vi-app}
For $p\in K$ and $q\in\R^d$,
\[
    p=\Proj_K(q)
    \quad\Longleftrightarrow\quad
    \langle q-p,w-p\rangle\le0
    \qquad \forall w\in K.
\]
\end{lemma}

\begin{proof}
The function $F(w)=\frac12\|q-w\|_2^2$ is convex and differentiable. A point $p\in K$ minimizes $F$ over $K$ if and only if every feasible directional derivative from $p$ is nonnegative:
\[
    \left.\frac{d}{dt}\right|_{t=0^+}F((1-t)p+tw)\ge0
    \qquad \forall w\in K.
\]
Computing the derivative gives $-\langle q-p,w-p\rangle\ge0$, proving the claim.
\end{proof}

Taking $q=z+\eta g$ and $p=z$ gives the projected stationarity condition
\[
    z=\Proj_K(z+\eta g)
    \quad\Longleftrightarrow\quad
    \langle g,w-z\rangle\le0
    \qquad \forall w\in K.
\]
Equivalently, $g\in N_K(z)$, where
\[
    N_K(z)=\{v:\langle v,w-z\rangle\le0 \text{ for all } w\in K\}
\]
is the outward normal cone. Thus interior projected stationarity requires $g=0$, while boundary stationarity allows outward directions.

For the interval $[a,b]$, the normal cone is
\[
N_{[a,b]}(s)=
\begin{cases}
(-\infty,0], & s=a,\\
\{0\}, & a<s<b,\\
[0,\infty), & s=b.
\end{cases}
\]
Hence the scalar projected condition is equivalent to the box sign rule:
\[
    s=a\Rightarrow g\le0,\qquad
    a<s<b\Rightarrow g=0,\qquad
    s=b\Rightarrow g\ge0.
\]

\begin{repeatlemma}{lem:box-rule}
Let $s\in[a,b]$, $g\in\mathbb{R}$, and $\eta>0$. Then
\[
    s=\Proj_{[a,b]}(s+\eta g)
\]
if and only if
\begin{equation}
    s=a \Rightarrow g\le 0,
    \qquad
    a<s<b \Rightarrow g=0,
    \qquad
    s=b \Rightarrow g\ge 0.
\end{equation}
\end{repeatlemma}
\begin{proof}
If $a<s<b$, projection is locally the identity, so $s=\Proj_{[a,b]}(s+\eta g)$ forces $g=0$. If $s=a$, the projection returns $a$ exactly when $a+\eta g\le a$, equivalently $g\le0$. If $s=b$, the projection returns $b$ exactly when $b+\eta g\ge b$, equivalently $g\ge0$.
\end{proof}

This boundary sign rule is the mechanism that creates cycles in our examples. In the quadratic gadget, each edge of $\calC$ has one free coordinate with zero gradient, while the remaining coordinates are pinned at endpoints by correct normal-cone signs.

We also use the following nonexpansiveness property.

\begin{lemma}
\label{lem:projection-nonexpansive-app}
For any nonempty closed convex set $K\subseteq\R^d$,
\[
    \|\Proj_K(u)-\Proj_K(v)\|_2\le \|u-v\|_2.
\]
\end{lemma}

\begin{proof}
Let $p=\Proj_K(u)$ and $q=\Proj_K(v)$. By \Cref{lem:projection-vi-app},
\[
    \langle u-p,q-p\rangle\le0,
    \qquad
    \langle v-q,p-q\rangle\le0.
\]
Adding gives
\[
    \|p-q\|_2^2
    \le
    \langle u-v,p-q\rangle
    \le
    \|u-v\|_2\|p-q\|_2,
\]
and the result follows.
\end{proof}

Consequently, a uniform perturbation of at most $\delta$ in the players' own-action gradients changes the projected residual by at most $\eta\delta$ throughout the strategy box. This bound is used in \Cref{lem:residual-stability,thm:c1}.

\subsection{Homology and Cohomology}
\label{app:homology-primer}

We recall the basic homology and cohomology facts used throughout the paper, following \citet{hatcher2002algebraic}. Singular homology allows us to characterize the topology of the constructed FONE sets, while cohomology enters the barrier theorem through its contravariance. In particular, a continuous map $f:X\to Y$ induces
\[
f^*:H^q(Y;\bbF)\to H^q(X;\bbF),
\]
which is the direction required by the factorization argument in \Cref{thm:master}.

For a circle and an isolated point, the relevant homology groups are
\[
H_1(S^1;\Z)\cong\Z,
\qquad
H_1(\{p\};\Z)=0.
\]
By additivity over disjoint unions, the set $A=S^1\sqcup\{p\}$ satisfies
\[
H_1(A;\Z)
\cong
H_1(S^1;\Z)\oplus H_1(\{p\};\Z)
\cong\Z.
\]
Thus, the isolated point contributes nothing to first homology, and the nontrivial topology comes entirely from the loop.

For coefficients in a field $\bbF$, the universal coefficient theorem gives
\[
H^q(X;\bbF)
\cong
\operatorname{Hom}_{\bbF}(H_q(X;\bbF),\bbF).
\]
Hence nontrivial homology over $\bbF$ is equivalent to nontrivial cohomology over $\bbF$. In particular,
\[
H_1(A;\Z)\cong\Z
\quad\Longrightarrow\quad
H^1(A;\bbF)\cong\bbF.
\]

Homotopy invariance further implies that a deformation retract has the same homology and cohomology as the ambient space. In particular, a tubular neighborhood of a polygonal circle deformation retracts onto the circle and therefore has the same first cohomology.

Finally, the block-product construction uses the K\"unneth formula over a field,
\[
H_n(X\times Y;\bbF)
\cong
\bigoplus_{i+j=n}
H_i(X;\bbF)\otimes_{\bbF}H_j(Y;\bbF).
\]
Applying this to the $r$-dimensional torus gives
\[
H_r((S^1)^r;\bbF)
\cong
H_1(S^1;\bbF)^{\otimes r}
\cong\bbF.
\]

\subsection{\v{C}ech Cohomology}
\label{app:cech-primer}

The exact FONE sets constructed in this paper are compact polyhedra, whereas approximate-FONE sets and residual sublevel sets under perturbations may have more irregular geometry. We therefore formulate the barrier for these approximate sets using \v{C}ech cohomology, following \citet{spanier1966algebraic,bredon1997sheaf,mardevsic1982shape}.

For a compact set $A\subseteq\R^d$, \v{C}ech cohomology admits the neighborhood characterization
\[
\check H^q(A;\bbF)
=
\varinjlim_{U\supseteq A} H^q(U;\bbF),
\]
where $U$ ranges over open neighborhoods of $A$, with restriction maps as the neighborhoods shrink. Thus, a class in $\check H^q(A;\bbF)$ can be represented by a cohomology class on some neighborhood of $A$, and two such representatives define the same class if their restrictions agree on a sufficiently small neighborhood.

\begin{lemma}
\label{lem:neighborhood-killing-app}
Let $A\subseteq\R^d$ be compact. If every open neighborhood $U\supseteq A$ contains a smaller neighborhood $V$ of $A$ such that the restriction map
\[
    H^q(U;\bbF)\to H^q(V;\bbF)
\]
is zero, then $\check H^q(A;\bbF)=0$.
\end{lemma}

\begin{proof}
Every \v{C}ech class is represented by a cohomology class on some neighborhood $U$ of $A$. By assumption, the representative restricts to zero on a smaller neighborhood $V$. Hence every representative dies in the direct limit.
\end{proof}

For compact polyhedra, singular and \v{C}ech cohomology agree, so our exact FONE sets, including polygonal cycles and isolated points, can be analyzed using either theory. Approximate-FONE sets may have more irregular geometry, for which \v{C}ech cohomology is particularly useful because of its continuity over shrinking neighborhoods.

\subsection{Dynamical Systems}
\label{app:dynamics-prelim}

We use standard notions of Lyapunov stability, attraction, and global attractors, following \citet{bhatia2002stability,la1976stability}. The central dynamical assumption in \Cref{thm:master} is uniform attraction, which requires that for every open neighborhood $U\supseteq A$, there exists an iterate that maps the entire compact state space into $U$.

This uniformity is essential to our proof. For each neighborhood $U$, the argument requires a single iterate $N$ satisfying
\[
\Psi_N(K)\subseteq U
\]
simultaneously for all initial conditions. Pointwise convergence allows the required number of iterations to depend on the initial condition and therefore does not guarantee the uniform inclusion needed for the cohomological factorization argument.

\subsection{Semialgebraic Regularity}
\label{app:semialgebraic}

A subset of $\R^d$ is semialgebraic if it can be described by a finite Boolean combination of polynomial equalities and inequalities. Such sets are triangulable and locally contractible \citep{bochnak2013real}. For polynomial games on boxes, projection acts by clipping each coordinate, so the projected update is polynomial on each of finitely many clipping regions. Consequently, the squared projected residual is piecewise polynomial, and its sublevel sets, including the unperturbed approximate-FONE sets, are semialgebraic.

\section{Proofs and Details}
\label{app:details}

This section provides supplementary details and omitted proofs from the main text.

\subsection{The Master barrier Theorem and its Variants}
\label{app:barrier-details}

The proof of \Cref{thm:master} rests on a simple factorization argument. Let $K$ be compact and contractible, and let $A\subseteq K$ be compact. Suppose that continuous maps $T_t:K\to K$ fix $A$ pointwise and uniformly attract to $A$. For any open neighborhood $U\supseteq A$, uniform attraction provides an $N$ such that
\[
\Psi_N(K)\subseteq U,
\qquad
\Psi_N=T_{N-1}\circ\cdots\circ T_0.
\]
Since each $T_t$ fixes $A$ pointwise, the inclusion $A\hookrightarrow U$ factors as
\[
A\hookrightarrow K\xrightarrow{\Psi_N}U.
\]
By contravariance of \v{C}ech cohomology, the induced restriction map
\[
\check H^q(U;\bbF)\longrightarrow\check H^q(A;\bbF)
\]
therefore factors through $\check H^q(K;\bbF)=0$ for every $q\ge1$, since $K$ is contractible. Thus every neighborhood-level cohomology class restricts trivially to $A$. Passing to the neighborhood direct limit gives $\check H^q(A;\bbF)=0$ for all $q\ge1$, contradicting the hypothesis of \Cref{thm:master}.

The argument relies on uniform attraction to choose a single iterate $\Psi_N$ for all initial conditions. We note that pointwise convergence alone does not guarantee this factorization.

\paragraph{Stable attraction.}
The same barrier applies to stable global attraction. Suppose $T:K\to K$ is continuous, $K$ is compact, and $A$ is Lyapunov stable and pointwise attracting. Fix an open neighborhood $U\supseteq A$. Stability gives a neighborhood $V\supseteq A$ such that every trajectory starting in $V$ remains in $U$. For each $z\in K$, pointwise attraction gives $n_z$ with $T^{n_z}z\in V$. By continuity, a neighborhood of $z$ also enters $V$ at time $n_z$. Compactness gives a finite subcover, hence a uniform time after which every trajectory lies in $U$. Therefore stable pointwise attraction implies uniform attraction, and \Cref{thm:master} applies.

\paragraph{Strict Lyapunov certificates.}
A strict Lyapunov certificate also implies uniform attraction on compact $K$. Suppose $V:K\to\R_{\ge0}$ is continuous, $V^{-1}(0)=A$, and
\[
    V(Tz)<V(z)
    \qquad \forall z\notin A.
\]
For any neighborhood $U\supseteq A$,
\[
    a_U=\min_{z\in K\setminus U}V(z)>0.
\]
On the compact set $\{z:V(z)\ge a_U\}$, the continuous function $V(z)-V(Tz)$ has a positive minimum. Hence every orbit enters $\{V<a_U\}\subseteq U$ in a uniformly bounded number of steps. Thus such a Lyapunov certificate would imply a pointwise-fixed uniformly attracting target, which is ruled out whenever $\check H^q(A;\bbF)\ne0$.

\subsection{Finite-memory Algorithms}
\label{app:memory-app}

A deterministic finite-memory method can be written on an extended state space
\[
    E=K\times M,
\]
where $M$ stores auxiliary variables such as momentum, previous gradients, optimism buffers, or adaptive statistics. If $K$ and $M$ are compact and contractible, then $E$ is compact and contractible.

The barrier applies to any compact target $B\subseteq E$ with $\check H^q(B;\bbF)\ne0$ that is fixed pointwise and uniformly attracting. The usual consistent-memory target is
\[
    B=\FONE(G)\times\{m_\star\},
\]
where $m_\star$ is a distinguished memory state, such as zero momentum. Then
\[
    \check H^q(B;\bbF)
    \cong
    \check H^q(\FONE(G);\bbF),
\]
so compact contractible memory does not remove the barrier.

This target assumption is important. A method could evade the theorem by attracting to a larger contractible set in extended state space whose projection is $\FONE(G)$. Such a method is not a pointwise-stationary solver for the exact FONE set in the sense studied here, because its actual attracting object does not have the topology of $\FONE(G)$.

\begin{repeatcorollary}{cor:finite-memory-gradient}[Finite-memory gradient methods]
Let $G$ be a game with compact contractible strategy space $K$ and FONE target
$A=\FONE(G)$. Suppose that $
    \check H^q(A;\mathbb{F})\neq 0
$
for some $q\ge 1$. Consider any deterministic finite-memory method with compact
contractible memory $M$ and continuous update rule
\[
    (z_{t+1},m_{t+1}) = T_G(z_t,m_t),
    \qquad
    (z_t,m_t)\in K\times M .
\]
Let $m_\star\in M$ be a distinguished consistent-memory state. If
\[
    T_G(a,m_\star)=(a,m_\star)
    \qquad \text{for every } a\in A,
\]
then the target $A\times\{m_\star\}$ cannot be uniformly globally attracting in
$K\times M$.

In particular, projected GDA with finitely many auxiliary variables,
extragradient and optimistic variants represented as finite-memory methods, and
momentum methods cannot admit a universal Lyapunov-style convergence proof that
both fixes every FONE and uniformly attracts to $\FONE(G)\times\{m_\star\}$ on
such games.
\end{repeatcorollary}

\begin{proof}
The extended state space $K\times M$ is compact and contractible. The target
$A\times\{m_\star\}$ has the same \v{C}ech cohomology as $A$. Applying
\Cref{thm:master} to the autonomous map $T_G$ on $K\times M$ rules out uniform
global attraction to $A\times\{m_\star\}$. The same argument applies to
time-varying finite-memory updates by using the nonautonomous version of
\Cref{thm:master}.
\end{proof}

\subsection{Homology of the Constructions and Block Products}
\label{app:examples-homology}
\label{app:kunneth}

The warm-up FONE set and the quadratic projected-gradient gadget in \Cref{sec:warmup,sec:gadget} both contain a circle component and an isolated point. Up to homeomorphism, they have the form
\[
    S^1\sqcup\{p\}.
\]
Therefore
\[
    H_1(S^1\sqcup\{p\};\Z)
    \cong
    H_1(S^1;\Z)\oplus H_1(\{p\};\Z)
    \cong
    \Z.
\]
The isolated point is homologically irrelevant in degree one; the barrier is carried by the loop.

For the quadratic gadget, the six-edge cycle $\calC$ is a polygonal circle, so $\calC\cong S^1$. Since
\[
    \PStat(\nabla\phi)=\calC\cup\{p\},
\]
we obtain
\[
    H_1(\PStat(\nabla\phi);\Z)\cong\Z.
\]
The diagonal map $z\mapsto(z,z)$ is a homeomorphism onto its image, so the lifted zero-sum game has the same topology:
\[
    \FONE(u_1,u_2)
    =
    \{(z,z):z\in\calC\}\cup\{(p,p)\}.
\]
Thus $\check H^1(\FONE(u_1,u_2);\bbF)\ne0$ for every field $\bbF$.

The same calculation amplifies under products. Let $A=\calC\cup\{p\}$. Then
\[
    A^r
    =
    \bigsqcup_{S\subseteq\{1,\ldots,r\}}
    \left(\prod_{i\in S}\calC\right)
    \times
    \left(\prod_{i\notin S}\{p\}\right).
\]
The component corresponding to $S=\{1,\ldots,r\}$ is
\[
    \calC^r\cong (S^1)^r=\mathbb{T}^r.
\]
By the K\"unneth formula over a field,
\[
    H_r(\mathbb{T}^r;\bbF)
    \cong
    H_1(S^1;\bbF)^{\otimes r}
    \cong
    \bbF.
\]
Hence $H_r(A^r;\bbF)\ne0$ and $\check H^r(A^r;\bbF)\ne0$. This is the topological calculation behind \Cref{thm:block-product}.

\subsection{Missing Details and Proofs for Warm-up in \Cref{sec:warmup}}
\label{app:missing_forced}

The warm-up construction tries to create a circle of stationary points in the interior. The following elementary argument explains why an additional interior critical point is unavoidable.

Let $f\in C^1(D^2)$ and suppose $\nabla f=0$ on $\partial D^2$. Then the tangential derivative of $f$ along $\partial D^2$ is zero, so $f$ is constant on $\partial D^2$. If $f$ is not constant on $D^2$, then either its maximum or minimum over the disk differs from this boundary value. That extremum is attained in the interior, and Fermat's rule gives an interior critical point. 

This is why the quadratic construction moves the nontrivial cycle to the boundary of a cube. Boundary normal-cone signs can create a stationary loop without forcing the loop to be an unconstrained interior critical set.

\begin{repeatthm}{thm:warmup}
For the game \eqref{eq:warmup-payoff},
\[
    \FONE(u_1,u_2)
    =
    \{(x,y):x^2+y^2=1\}\ \cup\ \{(0,0)\}.
\]
In particular, $H_1(\FONE;\mathbb{Z})\cong \mathbb{Z}$.
\end{repeatthm}
\begin{proof}
A direct calculation yields
\begin{equation}
\label{eq:warmup-gradients}
\partial_x u_1(x,y)
=2x\, h(x,y)\bigl(3+y^2-3x^2\bigr),
\qquad
\partial_y u_2(x,y)
=-2y\, h(x,y)\bigl(1+3y^2-x^2\bigr).
\end{equation}

We show that no boundary point satisfies the box sign conditions. At $x=2$, we have $h=3+y^2>0$ and $3+y^2-3x^2=y^2-9<0$, hence $\partial_x u_1<0$, violating the upper-endpoint condition. At $x=-2$, the sign reverses and violates the lower-endpoint condition. Similarly, at $y=2$ we obtain $\partial_y u_2<0$, and at $y=-2$ we obtain $\partial_y u_2>0$, both contradictions. Hence every FONE point lies in the interior.

In turn, we know that interior FONE points satisfy $\nabla_x u_1=0$ and $\nabla_y u_2=0$, so \eqref{eq:warmup-gradients} reduces to
\[
    x\,h(3+y^2-3x^2)=0,
    \qquad
    y\,h(1+3y^2-x^2)=0.
\]

If $h=0$, we recover the unit circle. Assume $h\neq 0$. If $x=0$, then the second equation gives $y(1+3y^2)=0$, hence $y=0$. If $y=0$, then the first equation gives $x(3-3x^2)=0$; the solutions $x=\pm1$ lie on $h=0$, so off the circle only $x=0$ remains. If $x,y\neq 0$, then
\[
    3+y^2-3x^2=0,
    \qquad
    1+3y^2-x^2=0.
\]
Substituting $x^2=1+3y^2$ into the first equation yields $-8y^2=0$, a contradiction. Hence the only off-circle solution is $(0,0)$.

Thus, the FONE set is the disjoint union of a circle and a point, so $H_1(\FONE;\mathbb{Z})\cong \mathbb{Z}$.
\end{proof}

\begin{figure}
\centering
\begin{tikzpicture}[scale=.72, line cap=round, line join=round]
  \coordinate (000) at (0,0);
  \coordinate (100) at (3,0);
  \coordinate (110) at (3,3);
  \coordinate (010) at (0,3);

  \coordinate (001) at (0.9,0.8);
  \coordinate (101) at (3.9,0.8);
  \coordinate (111) at (3.9,3.8);
  \coordinate (011) at (0.9,3.8);

  \draw[gray!55, thin] (001)--(101)--(111)--(011)--cycle;
  \draw[gray!55, thin] (000)--(001);
  \draw[gray!55, thin] (100)--(101);
  \draw[gray!55, thin] (110)--(111);
  \draw[gray!55, thin] (010)--(011);

  \draw[gray!55, thin] (000)--(100)--(110)--(010)--cycle;

  \foreach \v in {000,100,110,010,001,101,111,011} {
    \fill[red] (\v) circle (2pt);
  }

  \node[below left] at (000) {\scriptsize $(0,0,0)$};
  \node[below] at (100) {\scriptsize $(1,0,0)$};
  \node[right] at (110) {\scriptsize $(1,1,0)$};
  \node[right] at (111) {\scriptsize $(1,1,1)$};
  \node[above] at (011) {\scriptsize $(0,1,1)$};
  \node[left] at (001) {\scriptsize $(0,0,1)$};
  \node[left] at (010) {\scriptsize $(0,1,0)$};
  \node[right] at (101) {\scriptsize $(1,0,1)$};
\end{tikzpicture}
\caption{The projected stationary set of $\psi(x)=\frac12\|x\|^2$ on $[0,1]^3$. In particular, we have $\operatorname{PStat}(\nabla \psi)=\{0,1\}^3$, and hence $H_1(\PStat(\nabla \psi);\Z)=0$.}
\label{fig:pstat-positive-norm}
\end{figure}
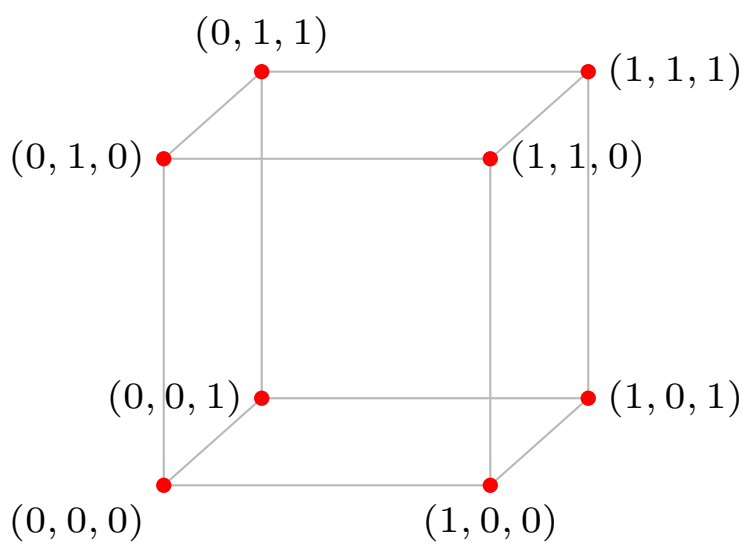

\subsection{Missing Details and Proofs for Quadratic Zero-sum Game in \Cref{sec:quadratic}}
\label{app:missing_diagonal}

\begin{lemma}
\label{lem:diagonal}
For the game \eqref{eq:main-game}, every FONE point satisfies $x=y$. On the diagonal, player~1's FONE condition reduces to $x\in\PStat(\nabla\phi)$. Consequently,
\[
    \FONE(u_1,u_2)=\{(z,z):z\in\PStat(\nabla\phi)\}.
\]
\end{lemma}

\begin{proof}
The own-gradients are
\[
    \nabla_x u_1(x,y)=\nabla\phi(x)+(x-y),
    \qquad
    \nabla_y u_2(x,y)=x-y.
\]
Fix a coordinate $i$. Player~2's own-gradient in coordinate $y_i$ is $x_i-y_i$. If $0<y_i<1$, the interior condition gives $x_i=y_i$. If $y_i=0$, the lower-endpoint condition gives $x_i\le 0$, and since $x_i\in[0,1]$, this forces $x_i=0=y_i$. If $y_i=1$, the upper-endpoint condition gives $x_i\ge 1$, and since $x_i\le 1$, this forces $x_i=1=y_i$. Hence $x=y$. On the diagonal, $\nabla_x u_1(x,x)=\nabla\phi(x)$ and $\nabla_y u_2(x,x)=0$, yielding the claim.
\end{proof}

\begin{repeatthm}{thm:main}[FONE set of our quadratic zero-sum game]
For the quadratic zero-sum game \eqref{eq:main-game},
\[
    \FONE(u_1,u_2)=\{(z,z):z\in\calC\}\cup\{(p,p)\}.
\]
Consequently $H_1(\FONE;\Z)\cong\Z$ and, by the universal coefficient theorem, $H^1(\FONE;\bbF)\cong\bbF$ for every field $\bbF$.
\end{repeatthm}
\begin{proof}
Combine \Cref{lem:diagonal} with \Cref{lem:gadget}. The diagonal map $z\mapsto(z,z)$ is a homeomorphism from $\calC\cup\{p\}$ onto $\FONE(u_1,u_2)$, so the homology statement follows. Since the FONE set is a compact polyhedron, singular and \v{C}ech cohomology agree, and the cohomology statement follows from the universal coefficient theorem over a field.
\end{proof}

\begin{repeatcorollary}{cor:main-impossibility}[Impossibility for quadratic games]
For the quadratic zero-sum game \eqref{eq:main-game}, no pointwise-fixed, continuous, uniformly attracting dynamics of the type covered by \Cref{thm:master,cor:dynamics-variants} can globally learn the exact FONE set. In particular, no Lyapunov-style argument satisfying \Cref{cor:no-lyapunov} can establish convergence to all FONE points.
\end{repeatcorollary}

\begin{proof}
The state space $[0,1]^3\times[0,1]^3$ is compact and contractible, while \Cref{thm:main} gives a FONE target with nonzero first \v{C}ech cohomology. The claim follows directly from \Cref{thm:master} and \Cref{cor:no-lyapunov}, including finite-memory extensions.
\end{proof}

\begin{repeatprop}{prop:minimal}[Degree two is minimal]
For any two-player zero-sum game on a box whose payoff has total degree at most one, the FONE set is a product of faces of the box. In particular, it is contractible and has trivial first homology.
\end{repeatprop}

\begin{proof}
Let $u_1(x,y)=a\cdot x+b\cdot y+c$ be affine and $u_2=-u_1$. Player~1's own-gradient is the constant vector $a$, and player~2's own-gradient is the constant vector $-b$. On each coordinate, the box rule selects the upper endpoint if the gradient is positive, the lower endpoint if it is negative, and the entire interval if it is zero. Therefore, the FONE set is a Cartesian product of intervals and endpoints, i.e., a face product of the strategy box. Such a set is contractible.
\end{proof}

\subsection{Missing Details and Proofs for Approximate FONE and Robustness in \Cref{thm:approx}}
\label{app:approx-more}

Let
\[
    S=\Delta\calC=\{(z,z):z\in\calC\}
\]
be the diagonal cycle in the exact FONE set of the quadratic game. Let $T$ be a tubular neighborhood of $S$, let $B$ be a small ball around $(p,p)$ disjoint from $T$, and set
\[
    O=\operatorname{int}(T\cup B).
\]
The component of $O$ containing $S$ deformation retracts onto $S$.

\begin{lemma}[Cycle sandwich]
\label{lem:cycle-sandwich-app}
Let $A$ be compact and satisfy
\[
    S\subseteq A\subseteq O.
\]
Then $\check H^1(A;\bbF)\ne0$.
\end{lemma}

\begin{proof}
Choose a cohomology class on $O$ whose restriction to the tubular-neighborhood component of $S$ generates $H^1(S;\bbF)$, and whose restriction to the ball component is zero. Its restriction to $S$ is nonzero because the tubular neighborhood deformation retracts onto $S$.

Suppose, for contradiction, that its image in $\check H^1(A;\bbF)$ were zero. By the neighborhood definition of \v{C}ech cohomology, the class would restrict to zero on some open neighborhood $W$ with
\[
    A\subseteq W\subseteq O.
\]
But $S\subseteq A\subseteq W$, so the restriction to $S$ would also be zero, a contradiction. Therefore $\check H^1(A;\bbF)\ne0$.
\end{proof}

For the exact game $G_\star$, the residual vanishes exactly on $\FONE(G_\star)$. Since $K\setminus O$ is compact and contains no exact FONE points,
\[
    m_O=\min_{w\in K\setminus O}\calR^\star_\eta(w)>0.
\]
Thus, for every $0<\eps<m_O$,
\[
    S\subseteq \AFONE_\eps(G_\star)\subseteq O.
\]
By \Cref{lem:cycle-sandwich-app},
\[
    \check H^1(\AFONE_\eps(G_\star);\bbF)\ne0.
\]

For a perturbed game $\widetilde G$, suppose the own-gradients are $\delta$-close uniformly. By the residual stability estimate, the residuals are $\eta\delta$-close. Writing $\tau=\eta\delta$, the same sandwich holds whenever
\[
    \tau\le \eps < m_O-\tau.
\]
Indeed, $S$ is contained in $\AFONE_\eps(\widetilde G)$ because the perturbed residual on $S$ is at most $\tau$, and no point outside $O$ is $\eps$-approximate because the perturbed residual there is at least $m_O-\tau$. Hence the approximate barrier survives small $C^1$ perturbations.

Consequently, the phenomenon is not a pathological artifact of one construction. In any finite-dimensional quadratic coefficient family containing $G_\star$, a sufficiently small open neighborhood of $G_\star$ consists of games whose approximate-FONE targets retain nontrivial first \v{C}ech cohomology for a nonempty interval of tolerances. We formalize this result in the following.

\begin{repeatthm}{thm:approx}[Approximate FONE and one-cycle]
For every $0<\eps<m_O$,
\[
    \Delta\calC\subseteq \AFONE_\eps(G_\star)\subseteq O.
\]
Moreover, $\check H^1(\AFONE_\eps(G_\star);\bbF)\ne0$. Consequently, $\AFONE_\eps(G_\star)$ cannot be a pointwise-fixed uniformly attracting target on the contractible strategy cube.
\end{repeatthm}
\begin{proof}
Since the residual vanishes on $\Delta\calC$, we have $\Delta\calC\subseteq \AFONE_\eps(G_\star)$. By definition of $m_O$, any point outside $O$ has residual at least $m_O>\eps$, so $\AFONE_\eps(G_\star)\subseteq O$.

Thus $\AFONE_\eps(G_\star)$ contains the core cycle $\Delta\calC$ and is contained in its neighborhood $T\cup B$. The inclusion of $\Delta\calC$ into $T$ induces a nontrivial map on first cohomology because $T$ deformation retracts onto $\Delta\calC$. Hence $\check H^1(\AFONE_\eps(G_\star);\bbF)\ne0$. The dynamical conclusion follows from \Cref{thm:master}.
\end{proof}

We now show that the residual remains stable under perturbations to the game. 

\begin{lemma}
\label{lem:residual-stability}
Let $G$ and $\widetilde G$ be games on the same box. If
\[
    \sup_K\|\nabla_xu_1-\nabla_x\widetilde u_1\|_2\le\delta,
    \qquad
    \sup_K\|\nabla_yu_2-\nabla_y\widetilde u_2\|_2\le\delta,
\]
then $ \sup_K|\calR_\eta^G-\calR_\eta^{\widetilde G}|\le \eta\delta $.
\end{lemma}

\begin{proof}
Projection onto a closed convex set is 1-Lipschitz, so each player's projected displacement changes by at most $\eta\delta$. Taking the maximum over the two players yields the claim.
\end{proof}

\begin{repeatthm}{thm:c1}
Let $\widetilde G$ be a game whose own-gradients are $\delta$-close to those of $G_\star$, and let $\tau=\eta\delta$. If $0<\tau<m_O/4$ and
$ \tau\le\eps<m_O-\tau $, then $
\Delta\calC\subseteq\AFONE_\eps(\widetilde G)\subseteq O,
$ and $\check H^1(\AFONE_\eps(\widetilde G);\bbF)\ne0$. Hence these approximate targets cannot be pointwise-fixed uniformly attracting targets.
\end{repeatthm}

\begin{proof}
By \Cref{lem:residual-stability}, $\|\calR_\eta^{\widetilde G}-\calR_\eta^\star\|_\infty\le\tau$. On $\Delta\calC$, the residual is zero for $G_\star$, so $\calR_\eta^{\widetilde G}\le\tau\le\eps$. If $w\notin O$, then $\calR_\eta^\star(w)\ge m_O$, hence $\calR_\eta^{\widetilde G}(w)\ge m_O-\tau>\eps$. The cohomology argument is identical to \Cref{thm:approx}.
\end{proof}

\begin{repeatcorollary}{cor:open-coefficient}
Fix any finite-dimensional parameterization of quadratic zero-sum games containing $G_\star$. There exists an open neighborhood $\calU$ of $G_\star$ such that for every $\widetilde G\in\calU$, there is a nonempty interval $I_{\widetilde G}$ with
\[
    \check H^1(\AFONE_\eps(\widetilde G);\bbF)\ne0
    \quad \text{for all } \eps\in I_{\widetilde G}.
\]
Consequently, sufficiently small approximate-FONE targets fail to be uniformly attracting on an open, positive-volume set of quadratic games.
\end{repeatcorollary}

\begin{proof}
Coefficient closeness implies uniform closeness of gradients on a compact domain. Choosing $\calU$ small enough ensures $\tau_{\widetilde G}<m_O/4$ for all $\widetilde G\in\calU$. The interval
\[
    I_{\widetilde G}=[\tau_{\widetilde G},\,m_O-\tau_{\widetilde G})
\]
is nonempty, and \Cref{thm:c1} applies. Openness implies positive Lebesgue measure.
\end{proof}

\newpage

\end{document}